\documentclass[preprint,12pt]{elsarticle}

\usepackage{amssymb, amsmath, amsfonts}
\usepackage{ulem}
\usepackage{amsthm}

\usepackage{graphicx}
\usepackage{textcomp}
\usepackage{array}
\newcolumntype{P}[1]{>{\centering\arraybackslash}p{#1}}
\newcolumntype{M}[1]{>{\centering\arraybackslash}m{#1}}
\usepackage{color}
\journal{Photoacoustics}

\begin{document}

\begin{frontmatter}



\title{Improving needle visibility in LED-based photoacoustic imaging using deep learning with semi-synthetic datasets}


\author[1]{Mengjie Shi}
\author[1]{Tianrui Zhao}
\author[2]{Simeon J. West}
\author[3,4]{Adrien E. Desjardins}
\author[1]{Tom Vercauteren}
\author[1]{Wenfeng Xia\corref{cor1}}
\ead{wenfeng.xia@kcl.ac.uk}
\cortext[cor1]{Corresponding author}
\address[1]{School of Biomedical Engineering and Imaging Sciences, King's College London, London SE1 7EH, United Kingdom}
\address[2]{Department of Anaesthesia,University College Hospital, London NW1 2BU, United Kingdom}         
\address[3]{Wellcome/EPSRC Centre for Interventional and Surgical Sciences, University College London, London W1W 7TY, United Kingdom}
\address[4]{Department of Medical Physics and Biomedical Engineering, University College London, London WC1E 6BT, United Kingdom}

\begin{abstract}
Photoacoustic imaging has shown great potential for guiding minimally invasive procedures by accurate identification of critical tissue targets and invasive medical devices (such as metallic needles). The use of light emitting diodes (LEDs) as the excitation light sources accelerates its clinical translation owing to its high affordability and portability. However, needle visibility in LED-based photoacoustic imaging is compromised primarily due to its low optical fluence. In this work, we propose a deep learning framework based on U-Net to improve the visibility of clinical metallic needles with a LED-based photoacoustic and ultrasound imaging system. To address the complexity of capturing ground truth for real data and the poor realism of purely simulated data, this framework included the generation of semi-synthetic training datasets combining both simulated data to represent features from the needles and \textit{in vivo} measurements for tissue background. Evaluation of the trained neural network was performed with needle insertions into blood-vessel-mimicking phantoms, pork joint tissue \textit{ex vivo} and measurements on human volunteers. This deep learning-based framework substantially improved the needle visibility in photoacoustic imaging \textit{in vivo} compared to conventional reconstruction by suppressing background noise and image artefacts, achieving 5.8 and 4.5 times improvements in signal-to-noise ratio (SNR) and the modified Hausdorff distance (MHD) respectively. Thus, the proposed framework could be helpful for reducing complications during percutaneous needle insertions by accurate identification of clinical needles in photoacoustic imaging. 
\end{abstract}



\begin{keyword}


Photoacoustic imaging\sep needle visibility \sep light emitting diodes \sep deep learning \sep minimally invasive procedures
\end{keyword}

\end{frontmatter}


\section{Introduction}
\label{Introduction}
Ultrasound (US) imaging is widely used for guiding minimally invasive percutaneous procedures such as peripheral nerve blocks \cite{chin2008needle}, tumour biopsy \cite{helbich2004stereotactic} and fetal blood sampling \cite{daffos1985fetal}. During these procedures, a metallic needle is inserted percutaneously into the body towards the target under real-time US guidance. Accurate and efficient identification of the target and the needle are of paramount importance to ensure the efficacy and safety of the procedure. Despite a number of prominent advantages associated with US imaging such as its real-time imaging capability, high affordability and accessibility, it suffers from intrinsically low soft tissue contrast that sometimes results in insufficient visibility of critical tissue structures such as nerves and small blood vessels. Moreover, visibility of clinical needles with US imaging is strongly dependent on the insertion angle and depth of the needle. With steep insertion angles, US reflections can be readily outside the transducer aperture, leading to poor needle visibility. Loss of visibility of tissue targets or the needle can cause significant complications \cite{rathmell2015safeguards}. 

Various methods have been developed for enhancing needle visualisation with US imaging, including echogenic needles \cite{hovgesen2022echogenica}, Doppler imaging \cite{fronheiser2008vibrating}, electromagnetic tracking \cite{klein2007piezoelectric}, and ultrasonic needle tracking \cite{xia2015inplane, xia2017looking, xia2017ultrasonic}. Although promising results have been reported, these methods usually require specialised equipment. Image-based needle tracking algorithms leveraging linear features of needles in US images have also been investigated such as random sample consensus \cite{uhercik2010model, waine20153d}, Hough Transform \cite{ding2003realtime, okazawa2006methods}, line filtering \cite{kaya2014needle, uhercik2013line}, and graph cut \cite{ayvaci2011biopsy}. However, it is challenging to automate these algorithms on US images with a variety of tissue backgrounds and needle contrasts. Deep learning (DL) based models especially convolutional neural networks have demonstrated competitive robustness and accuracy \cite{pourtaherian2018robust,arif2019automatic, gillies2020deep}, however, large clinical datasets with fine annotations are usually required for clinical applications but difficult to obtain.

Photoacoustic (PA) imaging has been of growing interest in the past two decades for its various potential preclinical and clinical applications, owing to its unique ability to resolve spectroscopic signatures of tissues at high spatial resolution and depths \cite{wang2016practical,ntziachristos2010going,beard2011biomedical}. In recent years, several research groups have proposed the combination of US and PA imaging for guiding minimally invasive procedures by offering complementary information to each other, with US imaging providing tissue structural information and PA imaging identifying critical tissue structures and invasive surgical devices such as metallic needles \cite{kuniyilajithsingh2016handheld,park2017realtime,xia2015performance,zhao2019minimally}. Recently, laser diodes (LDs) and light emitting diodes (LEDs) have shown promising results as an alternative to solid-stated lasers that are commonly used as PA excitation sources due to their favourable portability and affordability, which is of advantage to clinical translation \cite{xia2018handheld,kuniyilajithsingh2020portable,joseph_technical_2021}.

DL has been demonstrated as a powerful tool for signal and image processing, leading to remarkable successes in medical imaging \cite{lundervold2019overview, liu2019deep, domingues2020usinga, grohl2021deep}. DL-based approaches have been proposed for PA imaging enhancement, especially photoacoustic tomography \cite{hauptmann2020deep}, where they process raw channel data for image reconstruction \cite{antholzer2018photoacoustic, davoudi2019deep, lan2021deep} and enhancement \cite{allman2018photoacoustic, awasthi2020deep} as well as reconstructed images for image segmentation or classification \cite{boink2020partiallylearned, allman2019deep}.

DL has been recently used by several research groups for improving the imaging quality of LED-based PA/US imaging systems. Anas \textit{et al.} \cite{anas2018enabling}  exploited the use of a combination of a convolutional neural network (CNN) and a recurrent neural network (RNN) to enhance the quality of PA images by leveraging both the spatial features and temporal information in repeated PA image acquisitions. Kuniyil Ajith Singh \textit{et al.} \cite{kuniyilajithsingh2020portable} proposed a U-Net model to improve the SNR by training a neural network using PA images acquired by an improved PA imaging system with a higher laser energy and broadband ultrasound transducers. The pre-trained model was proven effective with LED-based PA images acquired from phantoms. Hariri \textit{et al.} \cite{hariri2020deep} proposed a multi-level wavelet-CNN to enhance noisy PA images associated with low fluence LED illumination by learning from PA images acquired with high fluence illumination sources. Enhancements were achieved on unseen \textit{in vivo} data with improved image contrast. Most recently, Kalloor Joseph \textit{et al.} \cite{kalloorjoseph2021generative} developed a generative adversarial network (GAN)-based framework for PA image reconstruction to mitigate the impact of the limited aperture and bandwidth of the ultrasound transducer. The proposed model was trained on simulated images from artificial blood vessels and tested on \textit{in vivo} measurements of the human forearm. The proposed approach was able to remove artefacts caused by the limited bandwidth and detection view. 

Although prominent attention has been given to improving the visualisation of tissue structures, notably, not much effort has been made to improve the visualisation of invasive medical devices in PA imaging. In this work, we proposed a DL-based framework to enhance the visibility of clinical needles with PA imaging for guiding minimally invasive procedures. As clinical needles have relatively simple geometries whilst background biological tissues such as blood vessels are complex, as opposed to using purely synthetic data \cite{heimann2014realtime, unberath2019enabling, maneas2021deep,movshovitz-attias2016how}, a hybrid method was proposed for generating semi-synthetic datasets \cite{garcia-peraza-herrera2021image}. The DL model was trained and validated using such semi-synthetic datasets and blind to the test data obtained from tissue-mimicking phantoms, \textit{ex vivo} tissue and human fingers \textit{in vivo}. The applicability of the proposed model on diverse \textit{in vivo} image data was further assessed on PA video sequences and compared with the standard Hough Transform. To the best of our knowledge, this is the first work that exploits DL for improving needle visualisation with PA imaging as well as utilizes semi-synthetic datasets for DL in PA imaging. 
\section{Material and methods}
\label{material and methods}
\subsection{System description}
A commercial LED-based PA/US imaging system (AcousticX, CYBERDYNE INC, Tsukuba, Japan) was used for acquiring experimental data. Detailed description of it can be found elsewhere \cite{singh2020led}. Briefly, PA excitation is provided by two LED arrays that sandwich a linear array US probe at a fixed angle. Each array consists of four rows of LEDs with 36 elements of 1 mm $\times$ 1 mm on each row. The LED arrays can be driven at different pulse repetition frequencies from 1 kHz to 4 kHz, and the maximum pulse energy from each array is 200 \textmu J. The LED pulse duration is controllable between 30 ns to 150 ns. In this study, a pulse width of 70 ns at 850 nm was selected for optimal energy efficiency \cite{IJET19907, hariri2018characterization}. The illumination area formed by the LED arrays was approximately a rectangle (50 mm $\times$ 7 mm), resulting in an optical fluence of 0.11 mJ/cm$^{2}$ at the maximum pulse energy of 400 \textmu J. The US probe has 128 elements over a linear array distance of 38.4 mm, with each element having a pitch of 0.3 mm, a central frequency of 7 MHz, and a -6 dB fractional bandwidth of 80.9\%.

Radio-frequency (RF) data for PA and US imaging were collected simultaneously from 128 channels on the probe with sampling rates of 40 MHz and 20 MHz, respectively. Interleaved PA and US imaging can be performed in real-time with image reconstruction performed on a graphics processing unit (GPU). Meanwhile, a maximum of 1536 PA frames and 1536 US frames corresponding to a total duration of 20s could be saved in memory at one time, available for offline reconstruction.
\subsection{Semi-synthetic dataset generation}
\begin{figure*}[!ht]
\centerline{\includegraphics[width= \textwidth]{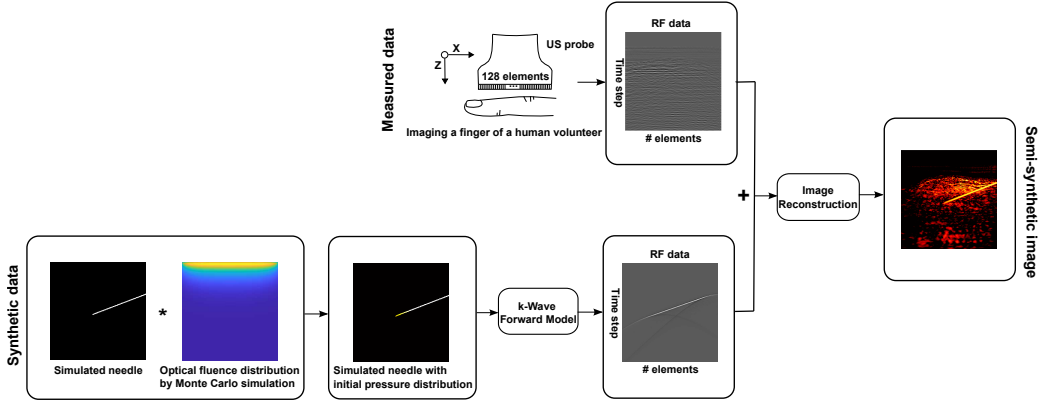}}
\caption{Flowchart illustration of the process of semi-synthetic training dataset generation. Top row: acquisition of sensor data from human finger vasculature \textit{in vivo} as background. Bottom row: synthetic radio-frequency (RF) sensor data generation from a simulated needle.}
\label{fig1-workflow}
\end{figure*}
The process of semi-synthetic dataset generation comprised three main steps as shown in Figure \ref{fig1-workflow}: 1) acquisition of \textit{in vivo} data to account for background photoacoustic signals originated from biological tissue; 2) generation of synthetic sensor data from needles; 3) image reconstruction with raw channel data combing synthetic and measurement data.

\textit{In vivo} data for background vasculature were collected by imaging the fingers of 13 healthy human volunteers using AcousticX. Experiments on human volunteers were approved by the King’s College London Research Ethics Committee (study reference: HR-18/19-8881). For each measurement, a total of 1536 PA frames and 1536 US frames was saved to the hard drive of the system's workstation for offline reconstruction. The RF data of one PA or US frame was a 2D matrix with a dimension of 1024 $\times$ 128. The first 150 (out of 1024) time steps were zeroed to remove strong LED-induced noise that spanned across the upper 5 mm depth in PA and US images. Averaging over 128 frames was implemented for suppressing random noise in the background. Reconstructed PA and US images corresponded to a field-of-view of 40.3 mm (X) $\times$ 39.4 mm (Z) according to the geometry of the linear transducer and the total number of time steps (1024) at a 40 MHz sampling rate.

Simulations of the sensor data originated from the needle were performed using the k-Wave toolbox \cite{treeby2010kwave}. Initial pressure distribution maps were created by simulating the optical fluence distributions on the needle shaft using Monte Carlo simulations \cite{wang1995mcml}. A 40.0 mm (X) $\times $ 40.0 mm (Z) region with a grid size of 0.1 mm was constructed to represent the background tissue. A uniform refractive index, optical scattering coefficient and anisotropy of 1.4, 10 mm$^{-1}$ and 0.9, respectively, were assigned to this region \cite{jacques2013optical}. Three optical absorption coefficients of 1 mm$^{-1}$, 1.5 mm$^{-1}$, 2 mm$^{-1}$ accounted for the variations in standard tissue. A homogeneous photo beam with a finite size of 38.4 mm was applied to the surface of the simulation area. Each simulation was run for around 10 minutes with approximate 100,000 photon packets. 

A linear array of 128 ultrasound transducer elements (with a pitch of 0.3 mm over a total length of 38.4 mm) were assigned to a forward model in k-Wave to receive the generated PA signals from the initial pressure distributions maps. The ultrasound transducer was assigned a central frequency of 7 MHz and a fractional -6 dB bandwidth of 80.9\% according to the specifications of AcousticX. RF data collected by the transducer elements were successively down-sampled to 40 MHz to match the sampling rate of the measured data. Considering the variations of needle insertions, simulations were conducted to account for clinically-relevant needle insertion depths and angles, spanning from 5 mm to 25 mm with an increment of 5 mm, and from 20 degrees to 65 degrees with a step of 5 degrees, respectively. 

To form a semi-synthetic image, the simulated RF data were normalised to maximum amplitude of \textit{ex vivo} needle signals collected by AcousticX. Subsequently, a pair of 2D data matrices (1024 $\times$ 128) consisted of RF data from a simulation on a needle and a measurement on a human finger were added to form a single 2D data matrix and then fed to a Fourier domain algorithm for image reconstruction \cite{jaeger2007fourier}. The reconstructed images based on the semi-synthetic data were then interpolated to 578 $\times$ 565 pixels with a uniform pixel size of 70 \textmu m $\times$ 70 \textmu m. To facilitate network implementations, the images were cropped to 512 $\times$ 512 pixels by removing the corresponding rows from top to bottom and the same number of columns from left to centre and right to centre respectively.

Finally, a total number of 2000 semi-synthetic images with substantial variations on both the needle and background were used for model training with the corresponding initial pressure distributions as the ground truths. 

\subsection{Acquisition of phantom, ex vivo and in vivo data for evaluation}
\label{data-acquisition}
Evaluation of the trained neural network was performed on PA images acquired with in-plane needle (20G, BD, USA) insertions into blood-vessel-mimicking phantoms, pork joint tissue \textit{ex vivo} and human fingers \textit{in vivo} (needle outside of tissue; see Supplementary Materials; Video S1). It is noted that the fingers with needle insertions were used to obtain representative real \textit{in vivo} data, but there is no corresponding clinical scenario.

The blood-vessel-mimicking phantoms were created by affixing several carbon fibre bundles in a plastic box filled with 1\% Intralipid dilution (Intralipid 20\% emulsion, Scientific Laboratory Supplies, UK) that had an estimated optical reduced scattering coefficient of 0.96 mm$^{-1}$ at 850 nm \cite{staveren1991light}. The fibre bundles were randomly positioned to mimic different orientations of blood vessels. The acquired PA images were prepared following the pipeline used for processing the semi-synthetic data.

\subsection{Network implementation}
\begin{figure}[!ht]
\centerline{\includegraphics[width=\columnwidth]{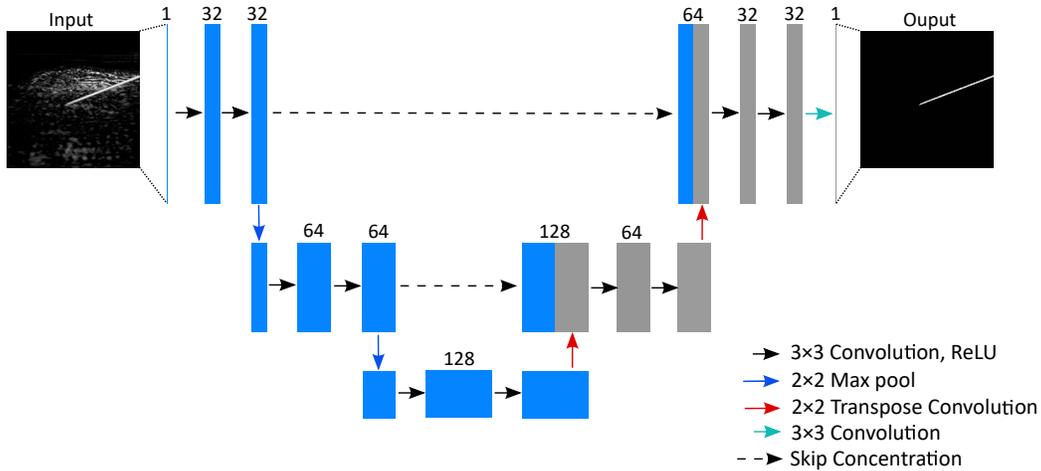}}
\caption{Architecture of the proposed network for improving needle visibility in photoacoustic imaging.}
\label{fig2-network}
\end{figure}

The network architecture implemented in this work was derived from the U-Net architecture proposed in Ref. \cite{hauptmann2020deep}. In general, this model followed the original U-Net architecture \cite{ronneberger2015unet} but had fewer scales and a reduced number of filters at each scale to accommodate a small input size. Experiments about the model capacity (see Supplementary Materials; Section 5) manifested that the model shown in Figure \ref{fig2-network} was able to learn the regularities of the training data and generalise well to unseen data. Besides, this model was built smaller and lighter which could contribute to size and latency reduction that are beneficial for real-time applications. 

In Figure \ref{fig2-network}, following an encoder path, each scale consisted of two convolutional layers followed by a 2 $\times$ 2 max pooling layer. For the decoder path, similarly, each scale contained two convolutional layers but followed by a transposed convolutional layer with an up-sampling factor of 2. The model was trained using the input pairs with a smaller resolution of 128 $\times$ 128 pixels that adapted well to the receptive field of the model by resizing from the initial size of 512 $\times$ 512 pixels via bicubic interpolation and evaluated on the real images with the size of 256 $\times$ 256 (see \ref{dc}. Discussion \& Conclusions regarding the choice of the image resolution).

Our network was implemented in Python using PyTorch v1.2.0. The semi-synthetic dataset was randomly split into training, test, and validation sets with a ratio of 8:1:1. Training was performed for 5000 iterations with a batch size of 4 that minimised the mean square error (MSE) loss in the validation set using the ADAM optimiser \cite{kingma2017adam} (initial learning rate: 0.001) and NVIDIA Tesla V100 GPUs. The CosineAnnealingLR learning rate scheme \cite{loshchilov2017sgdr} was employed to steadily decrease the learning rate during the training.

\subsection{Post-processing}
A post-processing algorithm based on maximum contour selection \cite{arbelaez2011contour} was employed for further improving the outcomes of the trained neural network and fitting the needle trajectory. It was assumed that for all the experiments only in-plane placements with one single needle was performed. The isolated outliers in the outputs of the U-Net could be discriminated regarding the region size different from the enhanced needle. Thus, the post-processing algorithm detected all contours in the outputs of the proposed model and saved the maximum contour as the one from the needle by counting the number of pixels on each contour boundary.

\subsection{Comparison method}
As a further evaluation step considering processing multiple PA frames from video sequences during dynamic needle insertions, the performance of the trained neural network on needle identification was compared with the standard Hough Transform (SHT), which is a classical baseline method for line detection \cite{duda1972use}. The SHT is designed to identify straight lines in images. It employs the parametric representation of a straight line, which is also called the Hesse normal form and can be expressed as: 
\begin{equation}
r = xsin\theta +ycos\theta
\end{equation}
where $r$ is the shortest distance from the origin to the line. $\theta$ measures the angle between the x-axis and the perpendicular projection from the origin point to the line. Therefore, a straight line can be associated with a pair of parameters ($r$, $\theta$), corresponding to a sinusoidal curve in Hough space. A few points on the same straight line will produce a set of sinusoidal curves that cross the same point ($r$, $\theta$) which exactly represents that line. In this study, the SHT was implemented by a two-dimensional matrix whose columns and rows were used to save the $r$ and $\theta$ values, respectively. For each point in the image, $r$ was calculated for each $\theta$, leading to increments of that bin in the matrix. Finally, the potential straight lines in the image were extracted by selecting the local maxima from the accumulator matrix.

\subsection{Evaluation protocol and metrics}
The needles in the acquired PA images from three different media (phantoms, pork joint tissue \textit{ex vivo}, and human fingers) were manually labelled as line segments by an experienced observer. The line needle segment was generated by connecting two points in the needle:  the needle tip and the farthest point to the tip on the needle shaft that was visualised in a PA image. The needle tip had a good contrast on US images when it was surrounded by water in the liquid-based tissue-mimicking phantoms, but had poor visibility for \textit{ex vivo} and \textit{in vivo} measurements. Therefore, solid glass spheres (0-63 \textmu m, Boud Minerals Limited, UK) were injected through the needle after being diluted with water to enhance the contrast of the tip, thus improving the accuracy and precision of manual labelling. For each medium, 20 representative measurements with different backgrounds and needle locations were used for metrics calculation.

To access the accuracy of needle extraction using the proposed DL-based approach, a metric called modified Hausdorff distance (MHD) was employed \cite{dubuisson1994modified}. The MHD was adapted from the generalised Hausdorff distances proposed for object matching with improved discriminatory power and greater robustness to outliers. Considering two point sets $\mathcal{A} = \{a_1, ..., a_{N_a}\}$ and $\mathcal{B} = \{b_1, ..., b_{N_b}\}$. The distance between a point $a$ from $\mathcal{A}$ and a set of points from $\mathcal{B}$ was defined as $d(a, \mathcal{B}) = min_{b\in\mathcal{B}}\Vert a-b\Vert$. The directed distance measure $d(\mathcal{A}, \mathcal{B})$ was defined as: 
\begin{equation}
    d(\mathcal{A}, \mathcal{B})=\frac{1}{N_a}\sum_{a\in\mathcal{A}}d(a, \mathcal{B})
\end{equation}
Then, the directed distance measures $d(\mathcal{A}, \mathcal{B})$ and $d(\mathcal{B}, \mathcal{A})$ were combined in the following way, resulting in the definition of the MHD as:
\begin{equation}
   \textup{MHD} = max(d(\mathcal{A}, \mathcal{B}), d(\mathcal{B}, \mathcal{A}))
\end{equation}
Signal-to-noise ratio (SNR) was also used to assess the performance of the needle enhancement and was defined as SNR = $\mathrm{S}$ / $\sigma$, where $\mathrm{S}$ is the mean amplitude over the needle region and $\sigma$ is the standard deviation of the background. The mean amplitude of the needle region was calculated by taking the average of the pixel values over the line segment. The background was defined as one of the largest rectangular regions that excluded the needle pixels.

\section{Results}
\subsection{Blood-vessel-mimicking phantoms}
\begin{figure*}[!ht]
\centerline{\includegraphics[width= \textwidth]{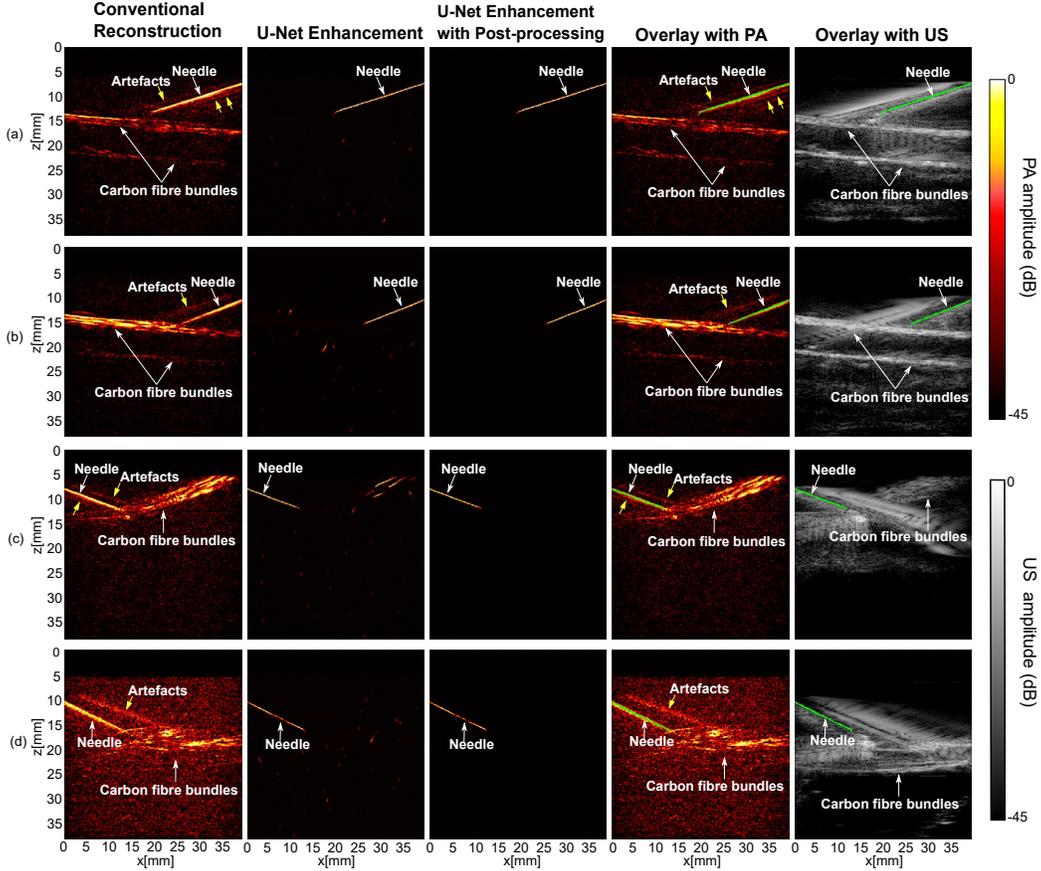}}
\caption{Photoacoustic imaging with needle insertions into a blood-vessel-mimicking phantom with conventional reconstruction, U-Net enhancement and U-Net enhancement with post-processing.}
\label{fig3-phantom}
\end{figure*}

The results of imaging on blood-vessel-mimicking phantoms are shown in Figure \ref{fig3-phantom}. Compared to conventional reconstructions, noticeable improvements in terms of removing background noise and artefacts can be observed in the outputs of the U-Net. The proposed model successfully identified the needle insertion without being perturbed by the background vessels that shared similar line-shape features. The false positives in the U-Net enhanced images were further suppressed by the post-processing. Besides, the proposed model demonstrated robustness to noise and strong artefacts (e.g., Figure \ref{fig3-phantom} (d)). The composite images indicated that the proposed model was able to detect the needle insertion with good correspondence to the conventional reconstructions and the US images. 

The performance of the proposed model was quantified with SNR and MHD (Table \ref{tab1-phantom}). Compared to the conventional reconstruction, the proposed model achieved a significant improvement in SNR by a factor of 8.3 ($p< .0001$). The MHD had large values as the noise level and artefacts increased. The proposed U-Net led to an initial 2.4 times decrease in MHD (from 63.2 $\pm$ 15.9 to 26.4 $\pm$ 23.3) and was further optimised by the successive post-processing to achieve the smallest MHD of 1.4 $\pm$ 1.3. 

\begin{table}[h]
\centering 
\caption{Quantitative evaluation of the trained neural network using blood-vessel-mimicking phantoms. These performance metrics are expressed as mean $\pm$ standard deviations from 20 measurements acquired from different phantoms and needle positions.}
\label{tab1-phantom}
\begin{tabular}{M{54pt}M{80pt}M{70pt}M{80pt}}
\hline 
Metrics&
Conventional Reconstruction&
U-Net Enhancement&
U-Net Enhancement with Post-processing \\
\hline 
SNR& 
$8.7 \pm 2.3 $&
$72.1 \pm 40.1 $&
- \\ 
\hline 
MHD& 
63.2 $\pm$ 15.9 &
$26.4 \pm 23.3 $&
$1.4 \pm 1.3$ \\ 
\hline 
\end{tabular}
\end{table}

\subsection{Pork joint tissue ex vivo}
\begin{figure*}[!ht]
\centerline{\includegraphics[width= \textwidth]{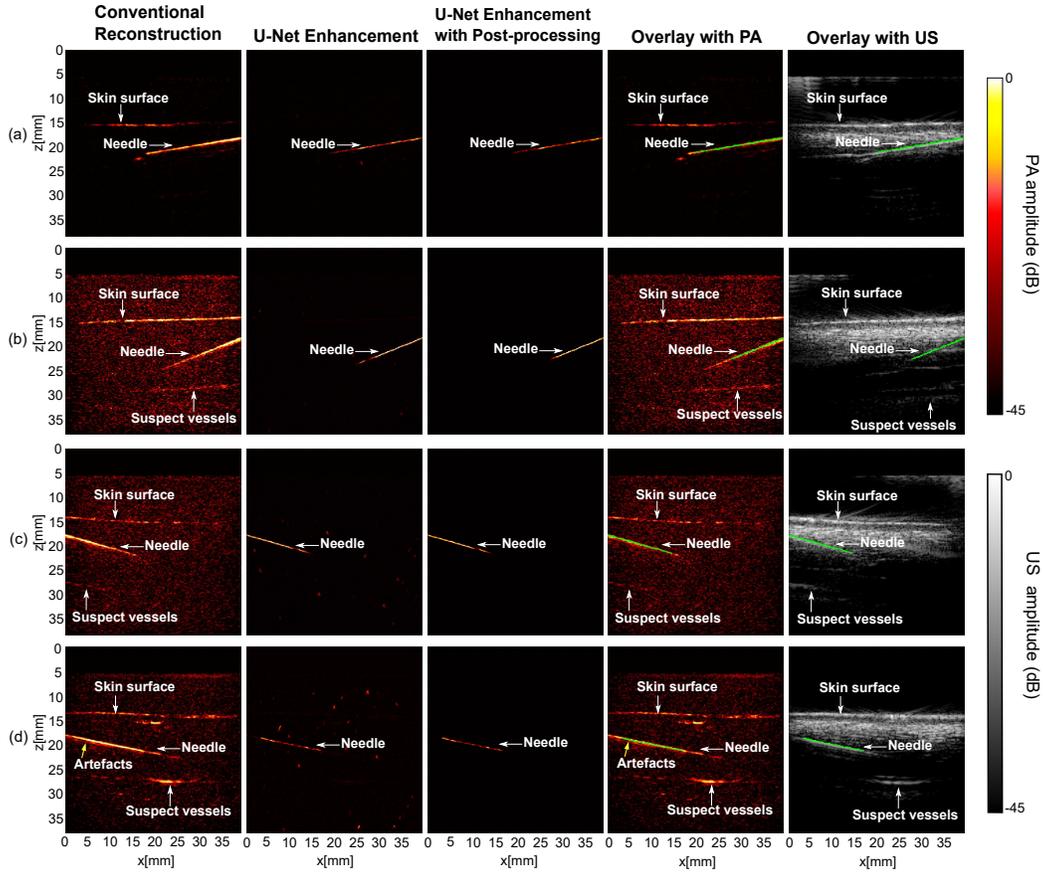}}
\caption{Photoacoustic imaging with needle insertions into \textit{ex vivo} tissue with conventional reconstruction, U-Net enhancement, and U-Net enhancement with post-processing.}
\label{fig4-pork}
\end{figure*}

The proposed model was also applied to images acquired from needle insertions into \textit{ex vivo} tissue. Figure \ref{fig4-pork} shows the representative results comparing the conventional reconstruction and the proposed model. The proposed model robustly enhanced the needle visibility with different insertion depths and angles, and significantly suppressed image artefacts and background noise. It also achieved a 4.8 times improvement in SNR compared to the conventional reconstruction ($p<.00001$)(Table \ref{tab2-pork-joint}). The MHD substantially decreased from 28.7 $\pm$ 16.3 to 4.5 $\pm$ 7.0 after post-processing the output of the U-Net ($p<.00001$).

It is worth noting that the performance of the proposed model was slightly degraded with visualising the region near the needle tip, which could be attributed to the large depths at around 2.5 cm (Figure \ref{fig4-pork} (b) and (d)). However, within smaller imaging depths, the proposed model was still effective for increasing the imaging speed by reducing the number of averages required to visualise the needle with a high SNR (see Supplementary Materials; Figure S4).

\begin{table}[h]
\centering 
\caption{Quantitative evaluation of the trained neural network using ex vivo needle images. These performance metrics are expressed as mean $\pm$ standard deviations from 20 measurements acquired from different spatial locations of the ex vivo tissue and needle positions.}
\label{tab2-pork-joint}
\begin{tabular}{M{54pt}M{80pt}M{70pt}M{80pt}}
\hline 
Metrics&
Conventional Reconstruction &
U-Net Enhancement&
U-Net Enhancement with Post-processing\\
\hline 
SNR& 
$19.0 \pm 9.7$&
$91.3 \pm 47.3$&
-\\ 
\hline 
MHD & 
28.7 $\pm$ 16.3 &
$6.3 \pm 9.1$&
$4.5\pm 7.0$\\ 
\hline 
\end{tabular}
\end{table}

\subsection{In vivo imaging}
\begin{figure*}[!ht]
\centerline{\includegraphics[width=\textwidth]{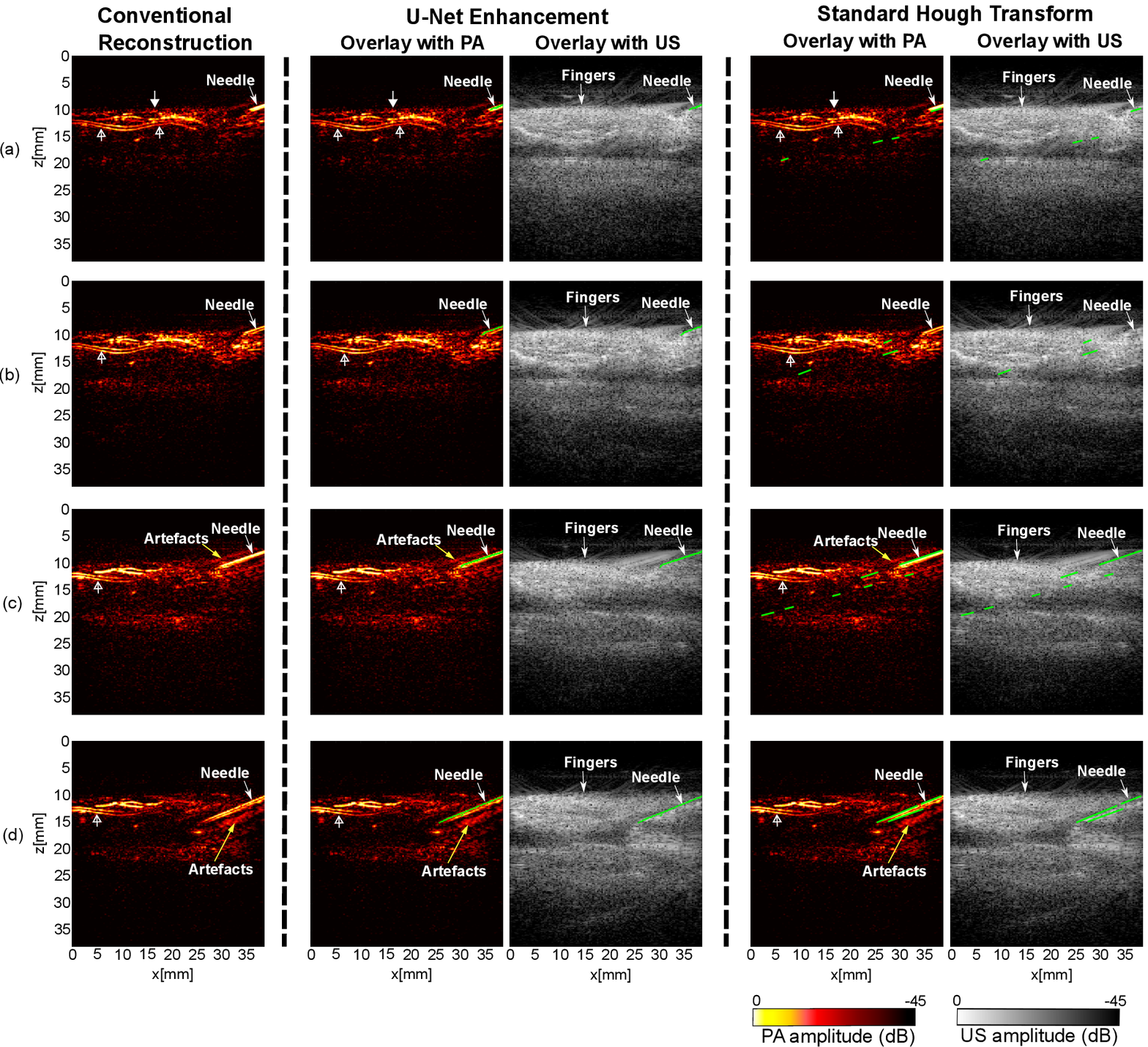}}
\caption{Photoacoustic (PA) imaging with needle insertions into human fingers \textit{in vivo} with conventional reconstruction, U-Net enhancement, and standard Hough Transform. Signals from the skin surface are indicated by triangle wide arrows, and signals that may be from digital arteries are indicated by hollow triangle wide arrows. The outcomes of U-Net enhancement and standard Hough Transform are denoted by green lines in PA and ultrasound (US) overlays. (a) - (d) are from a reconstructed PA image sequence recorded during needle insertions in real time.}
\label{fig7-invivo}
\end{figure*}

Figure \ref{fig7-invivo} shows the results of PA imaging with a 20G needle inserted between two fingers of a human volunteer immersed in a water tank (needle outside of the fingers). A main digital artery which has a two-layered feature is apparent on PA images (marked by hollow triangle wide arrows). A 22s video consisting of 128 frames was saved during the needle insertions (see Supplementary Materials; Video S2). Figure \ref{fig7-invivo} shows four frames acquired at different time points with the conventional reconstructions and the overlays of PA and US images after the U-Net enhancement and SHT. The SHT was able to detect the needle, but resulted in excessive false positives. This is because the performance of the SHT is sensitive to the specifications of hyperparameters, such as the detectable length of line segments and the searching resolution of $r$ and $\theta$. Fine-tuning of the hyperparameters based on the observations of the needle on a frame by frame basis is not trivial. In comparison, the proposed model manifested a good ability of robustness and generalisation. The \textit{in vivo} results demonstrated it's insensitive to constantly changing lengths and angles of the needle, and images with excessive noise and artefacts.

Quantitative results are summarised in Table \ref{tab3-finger}. An average of 5.8 times improvement in SNR was observed in the U-Net enhancement versus conventional reconstruction at different time points during the insertion ($p< .00001$). For MHD, the proposed model outperformed the conventional reconstruction and the SHT with a 4.5- and 2.9-fold reduction, respectively ($p< .00001$ for the conventional reconstruction; $p< .0001$ for the SHT). The post-processing algorithm effectively suppressed the outliers in the output of the U-Net, leading to the MHD as small as 0.6.

To further evaluate the model performance, needle detection rate was measured with three \textit{in vivo} PA sequences (Table \ref{tab4-sequences}; Supplementary Materials; Video S3). For each sequence containing 128 frames, the number of frames with the identifiable needle was counted by the observer as the reference for calculation. It is worth noting that the proposed model enhanced almost all the frames that contained the needle with a true positive rate up to 100\%, 90.6\%, and 97.0\% respectively.

\begin{table}[h]
\centering 
\caption{Quantitative evaluation of the trained neural network using in vivo needle images. These performance metrics are expressed as mean $\pm$ standard deviations from 20 measurements acquired at different time points during the insertion.}
\label{tab3-finger}
\begin{tabular}{M{54pt}M{55pt}M{70pt}M{80pt}M{55pt}}
\hline 
Metrics&
CR$^{a}$&
U-Net Enhancement&
U-Net Enhancement with Post-processing&
SHT$^{b}$\\
\hline 
SNR& 
$8.8 \pm 1.5  $&
$51.0 \pm 11.8 $&
-&
- \\ 
\hline 
MHD& 
87.7 $\pm$ 24.4 &
$19.4 \pm 5.0$&
$0.6 \pm 0.1$&
$55.7 \pm 38.5 $ \\ 
\hline 
\multicolumn{5}{r}{$^{a}$Conventional Reconstruction}{$^{b}$Standard Hough Transform}
\end{tabular}
\end{table}

\begin{table}[h]
\centering
\caption{Quantitative performance on three PA video sequences \textit{in vivo} with the proposed model}
\label{tab4-sequences}
\begin{tabular}{llll}
\hline
                                                                                     & Sequence 1 & Sequence 2 & Sequence 3 \\ \hline
\begin{tabular}[c]{@{}l@{}}Frames with needle\\ /Total frames\end{tabular} & 92/128     & 53/128     & 99/128     \\ \hline
\begin{tabular}[c]{@{}l@{}}Frames without needle\\ /Total frames\end{tabular}  & 36/128     & 75/128     & 29/128     \\ \hline
Needle missed                                                                        & 0          & 5          & 3          \\ \hline
True positives                                                                       & 92         & 48         & 96         \\ \hline
False positives                                                                      & 0          & 1          & 1          \\ \hline
True positive rate (\%)                                                              & 100        & 90.6       & 97.0       \\ \hline
False positive rate (\%)                                                             & 0          & 1.3        & 3.4        \\ \hline
\end{tabular}
\end{table}

\subsection{Impact of the needle diameter}
To further assess the generalisability of the proposed model, in-plane insertions of needles with different diameters (16G, 18G, 20G, BD, USA; 25G, 30G, Meso-relle, Italy) were imaged with AcousticX during in plane insertions into pork joint tissue \textit{ex vivo}. The results (see Supplementary Materials; Figure S5) were consistent with the previous results of the 20G needle. For the needles with small diameters, the acquired PA images readily suffered from lower SNRs. However, as expected, the proposed model yielded robust enhancement on the needles at different contrast levels. For SNR and MHD, the proposed model outperformed the conventional reconstruction with an average of 11.2 and 6.5 times improvement, respectively (see Supplementary Materials; Table S1). 

\section{Discussion}
\label{dc}
Previous works on DL in PA imaging mainly focused on improving the visualisation of the vasculature by denoising and artefacts removal. However, the performances of DL networks are highly dependent on the training dataset. Networks that are specifically trained to enhance the visualisation of vasculature usually have poor performance on visualising needles due to their different image features. In this work, we are the first to apply DL to specifically improve the needle visualisation with PA imaging for minimally invasive guidance. Considering the relatively simple geometries of clinical needles compared to vasculature, a prominent contribution of this work is that we developed a semi-synthetic approach to address the challenges associated with obtaining ground truth for \textit{in vivo} data as well as the poor realism of purely simulated data.

According to our experimental results (not shown for the sake of brevity), the simulated optical fluence distribution had a minimal effect on the performance of the proposed model when evaluated on unseen real needle images even with deep insertion angles. This is because the DL-based method was able to enhance the visualisation of the needle by learning its relatively simple spatial features that remained largely consistent. In addition, we found that the inference performance of the trained model on the real images did not benefit from a higher resolution input data than that is currently used (128 $\times$ 128 pixels; See Supplementary Materials; Figure S7). The low resolution images performed sufficiently well considering the lightweight model and simple features of the input data. Compared to a high resolution input, a low resolution input is also advantageous in terms of the computational costs; the inferring time for one image was around 90 ms using one GPU (NVIDIA GeForce RTX 2070 Max-Q). Further reduction on the inferring time could be realised by using more powerful GPUs for real-time applications.

During minimally invasive procedures, accurate and clear visualisation of the needle is essential for successful outcomes. Needle visibility has been greatly improved by PA imaging as compared to US imaging, but the image quality in terms of SNR with the LED light source is still sub-optimum due to the low pulse energy. Frame averaging is effective for reducing background noise, but at the cost of the imaging speed and introduces movement artefacts. Further, blood vessels in the background with similar line-shape structures to the needle are readily regarded as visual disturbances for clinicians to identify the needle trajectory. Finally, line artefacts above or beneath the needle shaft are often non-negligible that can lead to misinterpretation of the needle position. Therefore, in this work, PA images of needle insertions into different types of blood-vessel-mimicking phantoms, \textit{ex vivo} tissue, and \textit{in vivo} human fingers were acquired to evaluate the proposed model. 

Qualitative results demonstrated that our proposed model was able to achieve substantial enhancement on the needle visualisation regarding noise suppression, artefacts removal, and needle detection. The enhancement was further quantified by the SNR and MHD. Performance of the proposed model compared to the conventional reconstruction and the standard Hough transform on images acquired from blood-vessel-mimicking phantoms, \textit{ex vivo} pork joint tissue, and human fingers are shown in Supplementary Materials (Figure S1; Figure S2; Figure S3). For SNR, our proposed model achieved 8.3, 4.8, and 5.8 times enhancement for phantom, \textit{ex vivo}, and \textit{in vivo} data respectively. The MHD as a measure of similarity of two objects was employed for its great robustness and discriminatory power. It was observed that the MHD had the smallest values with our proposed model compared to the conventional reconstruction and the SHT (1.4, 4.5, and 0.6 for phantom, \textit{ex vivo}, and \textit{in vivo} data respectively). Additionally, it is evidenced that the post-processing method based on maximum contour selection was effective to remove the false positives of the U-Net enhancement while preserving the needle pixels.

We also compared the proposed model with a conventional line detection algorithm, SHT, with \textit{in vivo} video sequences. The SHT performed quite well on some cases with carefully chosen critical hyperparameters, but its performance was readily affected by imperfection errors from the former edge detection step and sensitive to some decision criteria such as empirical values of $r$ and $\theta$ that are directly related to the detection efficiency. Fine-tuning of these hyperparameters is impractical for real-time applications where the effective length and angle of the needle placements could constantly vary in each frame. In contrast, our proposed model can efficiently improve the needle visualisation on a variety of PA images from \textit{in vivo} measurements in near real-time.

Nonetheless, the DL-based enhancement was sensitive to the SNRs of the images. More importantly, the visibility of the needle, especially its tip was still limited to a depth of around 1 cm with \textit{in vivo} measurements. In the future, deep neural networks could be applied for real-time denoising \cite{hariri2020deep} as an alternative to frame-to-frame averaging to improve the imaging depth. For needle tip visualisation, a fibre-optic US transmitter could be integrated within the needle cannula so that the needle tip can be unambiguously visualised in PA imaging with high SNRs \cite{xia2017ultrasonic}. Light-absorbing coatings based on elastomeric nanocomposites could also be applied to the needle shaft for enhancing its visualisation for guiding minimally invasive procedures \cite{xia2019enhancing}.

\section{Conclusions}
\label{cl}
In this work, we provided a DL-based framework for enhancing needle visualisation with PA imaging. The DL-model was built using only semi-synthetic data generated by combining simulated data and \textit{in vivo} measurements. Evaluation was performed on unseen real data acquired by inserting needles into blood-vessels-mimicking phantoms, \textit{ex vivo} tissue and human fingers (needle outside tissue). Compared to the conventional reconstruction, the proposed framework substantially improved the needle visualisation with PA imaging. It also outperformed the standard Hough Transform on PA \textit{in vivo} videos with improved robustness and generalisability. Therefore, our framework could be useful for guiding minimally invasive procedures that involve percutaneous needle insertions by accurate identification of clinical needles. 
\section{Declaration of competing interests}
The authors declare that they have no known competing financial interests or personal relationships that could have appeared to influence the work reported in this paper. A.E.D. is a Director and Shareholder of Echopoint Medical, London, UK. T.V. is co-founder and shareholder of Hypervision Surgical Ltd, London, UK. He is also a shareholder of Mauna Kea Technologies, Paris, France.
\section{Acknowledgements}
This work was supported by the Wellcome Trust [203148/Z/16/Z, WT101957, 203145Z/16/Z], the Engineering and Physical Sciences Research Council (EPSRC) (NS/A000027/1, NS/A000050/1, NS/A000049/1), and King’s–China Scholarship Council PhD Scholarship Program (K-CSC) (202008060071).



\bibliographystyle{elsarticle-num} 
\bibliography{00}{}


%\begin{thebibliography}{00}


%\bibitem{}

%\end{thebibliography}

\end{document}

\title{Supplementary materials for ‘‘Improving needle visibility in LED-based photoacoustic imaging using deep learning with semi-synthetic datasets"}
\author{Mengjie Shi, Tianrui Zhao, Simeon J. West, Adrien E. Desjardins, \\Tom Vercauteren, and Wenfeng Xia}
\maketitle

\section{Metrics evaluation}
\begin{figure}[htb]
\centerline{\includegraphics[width=0.6\textwidth]{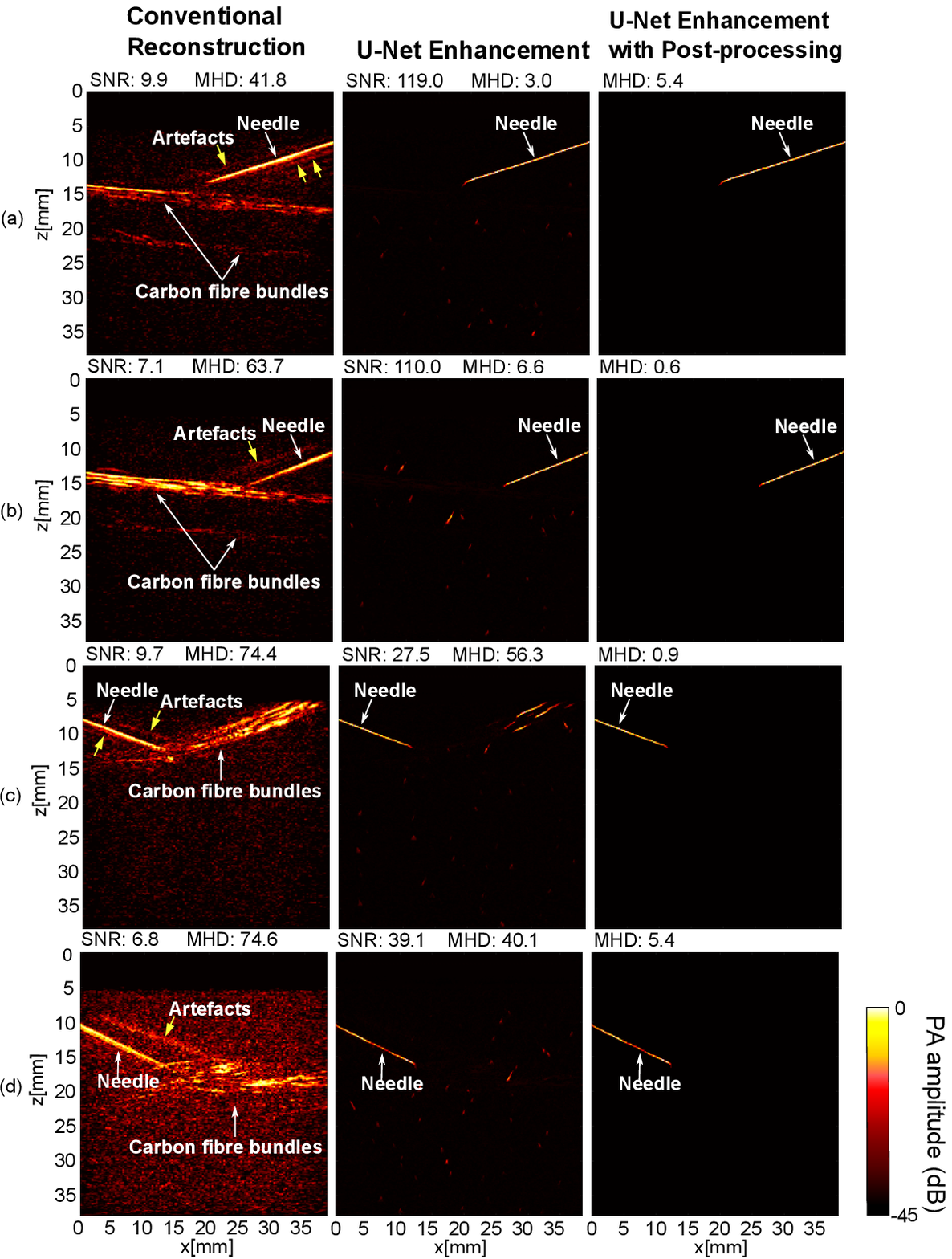}}
\caption{Photoacoustic imaging of needle insertions into blood-vessel-mimicking phantoms with conventional reconstruction, U-Net enhancement, and U-Net enhancement post-processing.}
\label{supp:snr-mhd-phantom}
\end{figure}

\begin{figure}[htb]
\centerline{\includegraphics[width=0.6\textwidth]{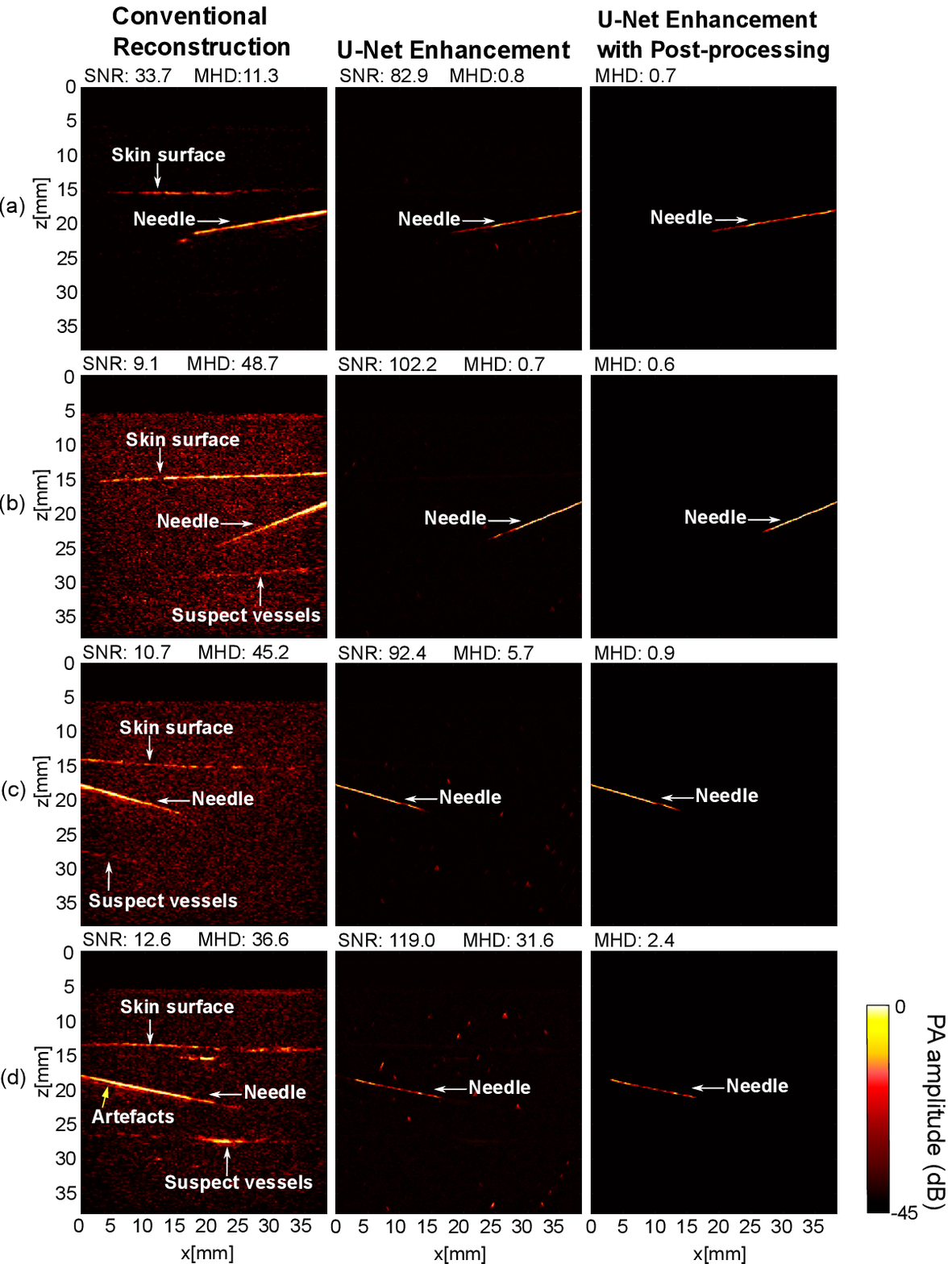}}
\caption{Photoacoustic imaging of needle insertions into pork joint \textit{ex vivo} with conventional reconstruction, U-Net enhancement, and U-Net enhancement post-processing.}
\label{supp:snr-mhd-pork}
\end{figure}

\begin{figure}[htb]
\centerline{\includegraphics[width=0.6\textwidth]{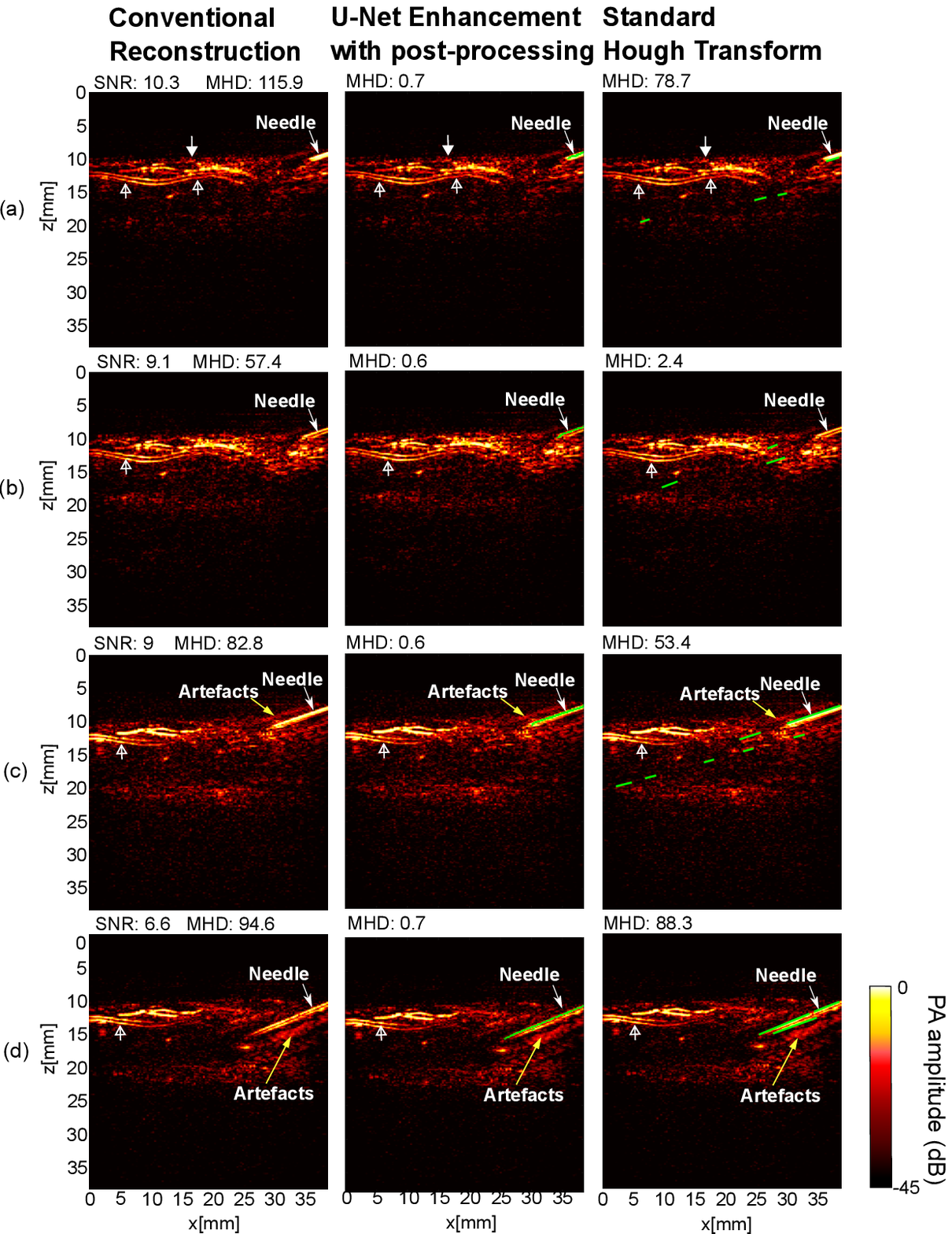}}
\caption{Photoacoustic imaging of needle insertions between two fingers of a human volunteer with conventional reconstruction, U-Net enhancement, and U-Net enhancement post-processing. The outcomes of the U-Net enhancement and the standard Hough Transform (SHT) are denoted by green lines in photoacoustic (PA) overlays. Signals that may correspond to skin and digital arteries are implied by triangle wide arrows and hollow triangle wide arrows respectively. }
\label{supp:snr-mhd-fingers}
\end{figure}

The signal-to-noise (SNR) and the modified Hausdorff distance (MHD) were employed for evaluating the model performance on phantom, \textit{ex vivo}, and \textit{in vivo} data. The quantitative results were reported for each representative image and shown in Figure \ref{supp:snr-mhd-phantom}, Figure \ref{supp:snr-mhd-pork}, and Figure \ref{supp:snr-mhd-fingers}.

\section{Impact of number of averaged frames}
\begin{figure}[htb]
\centerline{\includegraphics[width=\textwidth]{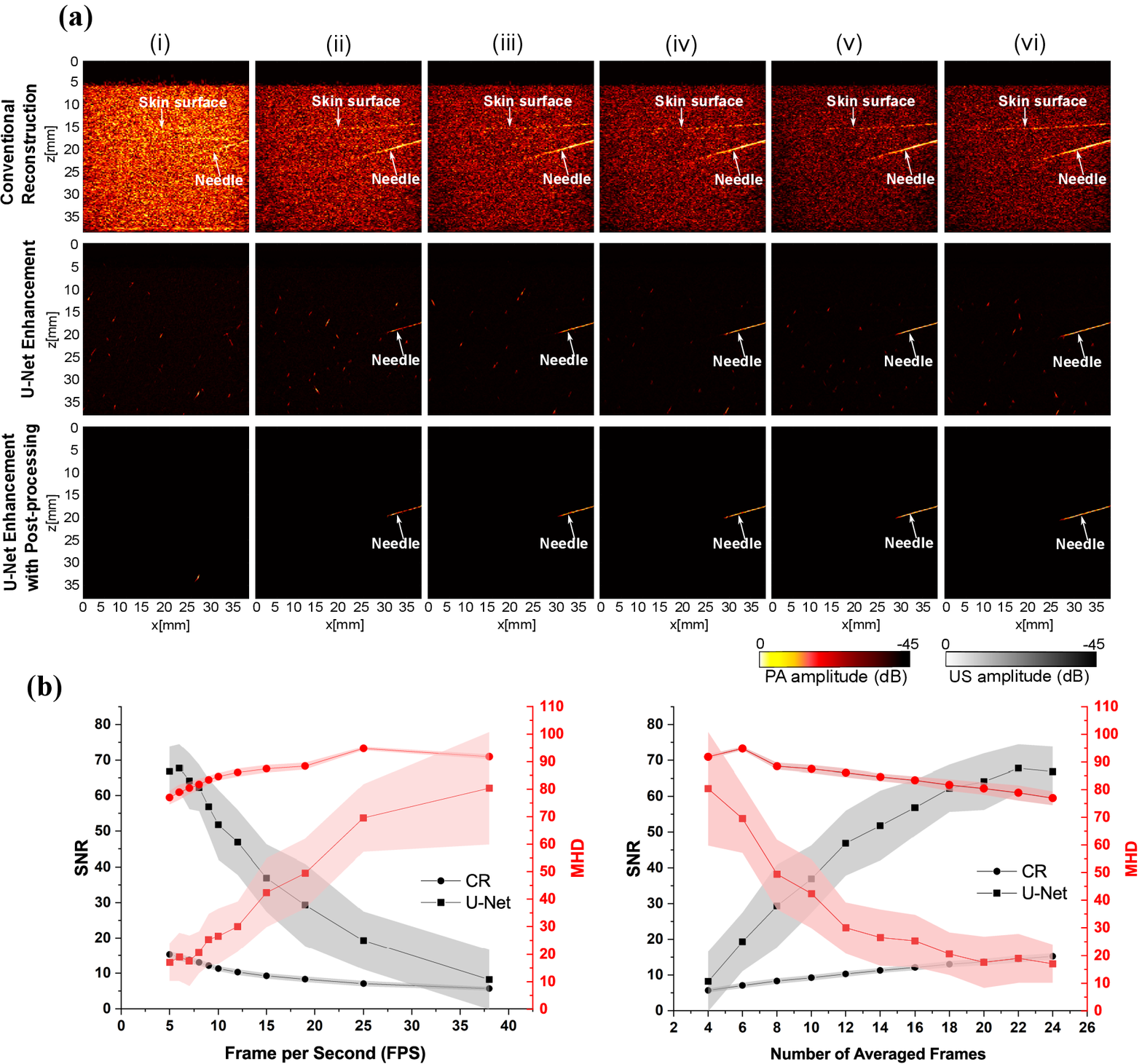}}
\caption{(a) Photoacoustic images of needle insertions into pork joint \textit{ex vivo} tissue with conventional reconstruction, U-Net enhancement and U-Net enhancement with post-processing under different number of averaged frames: from (i) to (vi) are 4, 8, 12, 16, 20, 24 respectively; (b) SNRs of \textit{ex vivo} needle images with conventional reconstruction (CR) and U-Net enhancement under different frames per second (FPS) (left) and different number of averaged frames (right). Data represent average values and shaded error bars represent standard deviations.}
\label{supp:results-porkjointaveraging}
\end{figure}

The relationship between the number of averaged frames (or imaging speed) and the SNRs and MHDs of the needle images with the U-Net enhancement and conventional reconstruction was investigated in Figure \ref{supp:results-porkjointaveraging}. The number of averaged frames was increased from 4 to 24 in 2 increments corresponding to an imaging speed decreased from 38 to 6 frames per second if only the data collection time was considered. Four groups of \textit{ex vivo} needle images (20 images for each group) with varying insertion depths and angles were used. 

\begin{figure}[htb]
\centerline{\includegraphics[width=\textwidth]{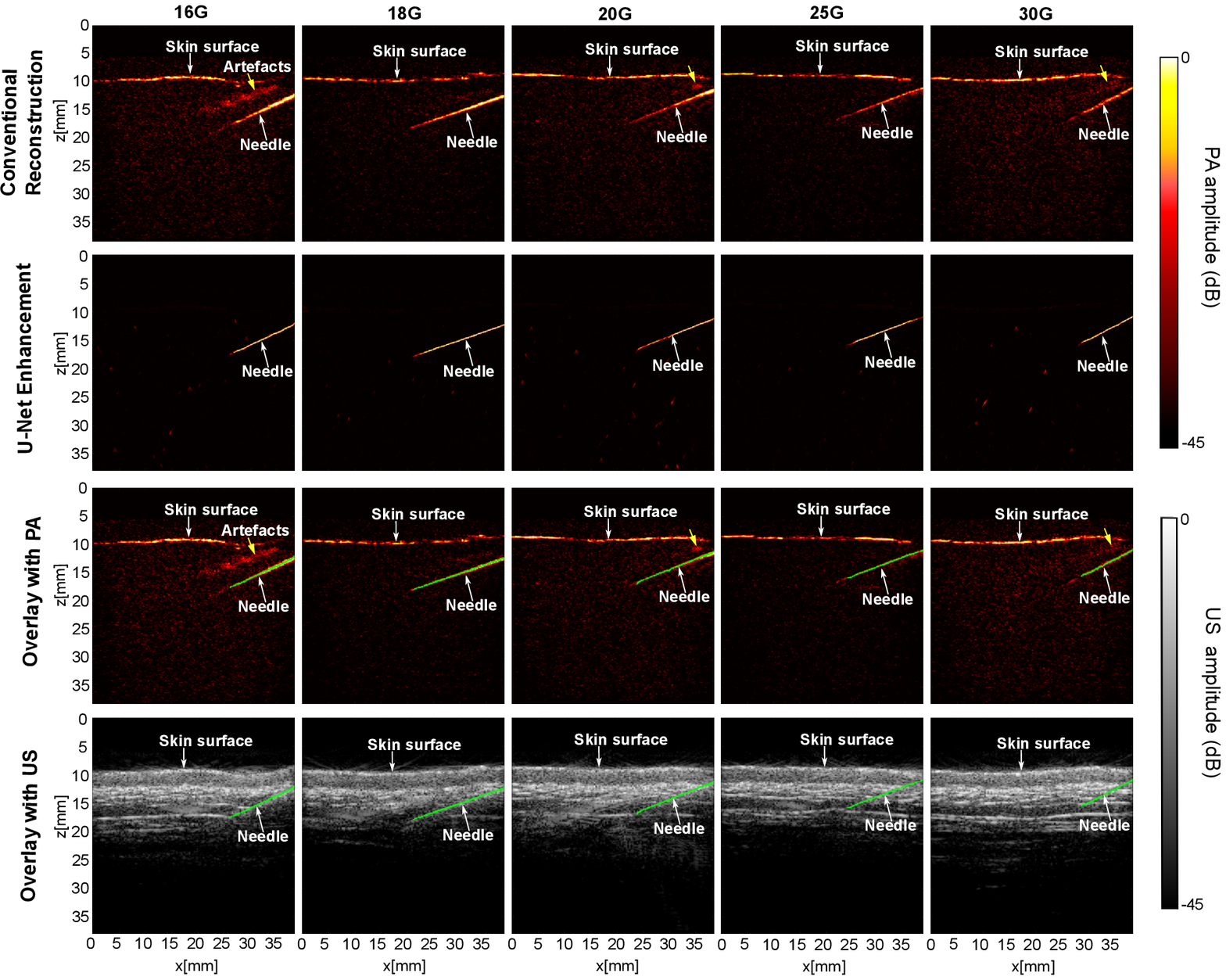}}
\caption{Photoacoustic imaging of needle insertions into pork joint \textit{ex vivo} tissue for needles with different diameters comparing conventional reconstruction, U-Net enhancement, and U-Net enhancement post-processing.}
\label{supp:results-needlesize}
\end{figure}

\section{Impact of needle diameters}
\begin{figure}[!ht]
\centerline{\includegraphics[width=0.5\textwidth]{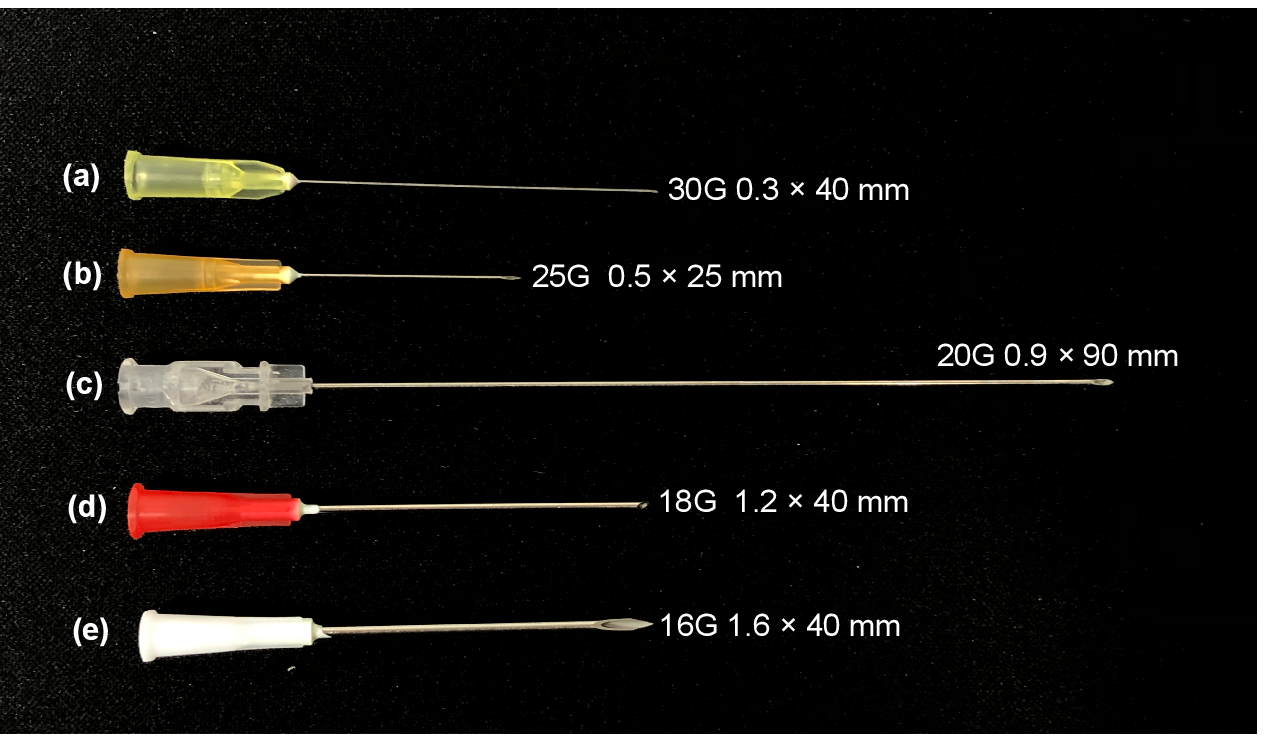}}
\caption{Photograph of metallic needles with different diameters: (a) 16G (b) 18G (c) 20G (d) 25G (e) 30G.}
\label{sup:needles}
\end{figure}

\section{Impact of input size}
\begin{figure}[htb]
\centerline{\includegraphics[width=\textwidth]{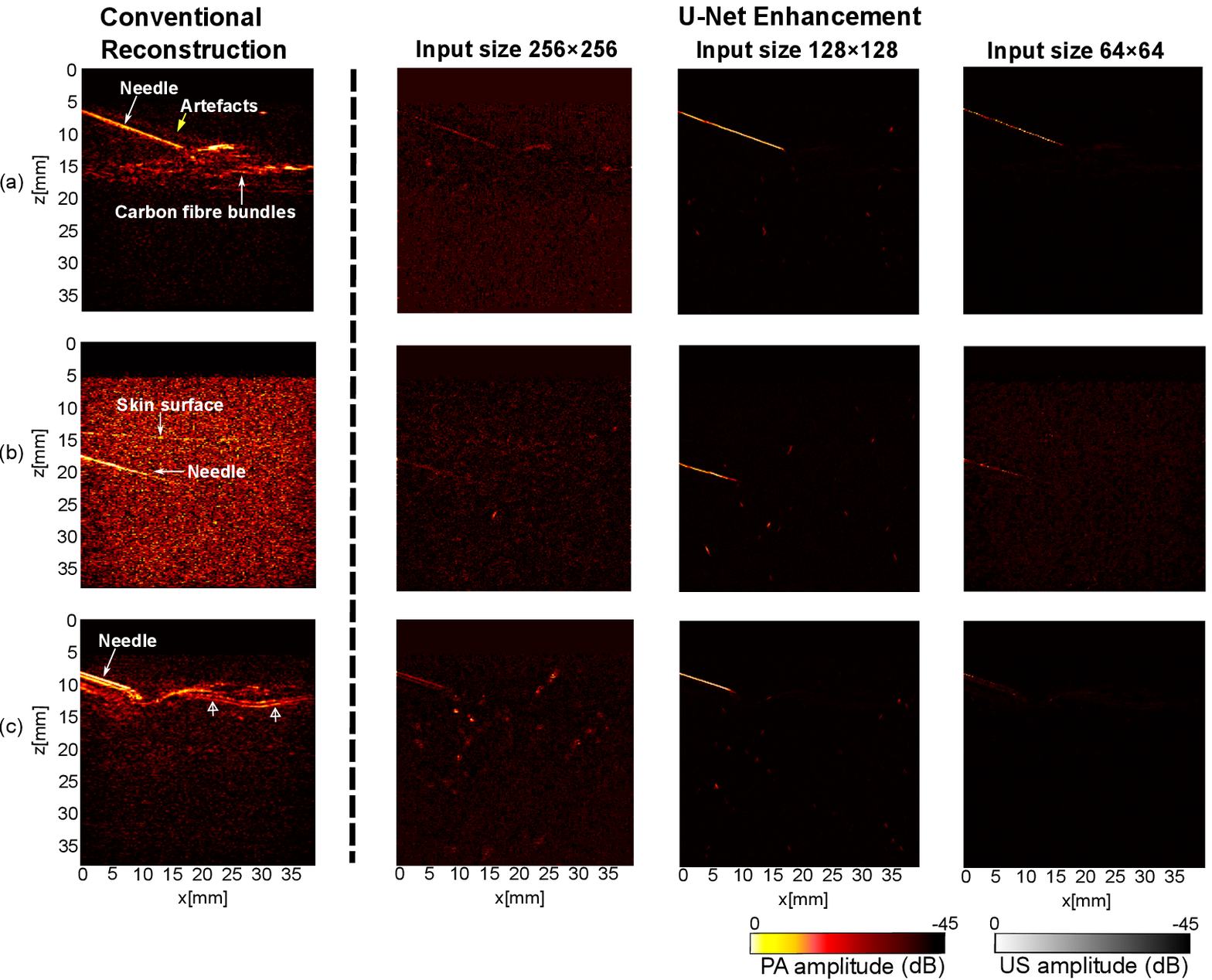}}
\caption{Model inference on different input sizes with real needle images from (a) blood-vessel-mimicking phantoms, (b) pork joint tissue \textit{ex vivo}, and (c) human fingers.}
\label{supp:results-diffres}
\end{figure}

\begin{table}[h]
\centering
\caption{Quantitative evaluation of the performance of trained neural network with respect to needles with different diameters using measurements on \textit{ex vivo} tissue. The performance metrics are expressed as mean $\pm$ standard deviations from 20 measurements acquired at different spatial locations of the \textit{ex vivo} tissue and needle positions.}
\label{tab4-gauges} 
\begin{tabular}{lll|lll}
\hline
                              & \multicolumn{1}{c}{SNR} &               & \multicolumn{3}{c}{MHD}                                                                                                 \\ \hline
\multicolumn{1}{c}{\textbf{}} & CR                      & U-Net         & CR            & U-Net        & \multicolumn{1}{c}{\begin{tabular}[c]{@{}c@{}}U-Net with\\ Post-processing\end{tabular}} \\ \hline
16G                           & 10.9$\pm$2.5            & 106$\pm$4.8   & 51.0$\pm$7.3  & 10.4$\pm$5.5 & 1.1$\pm$1.0                                                                              \\ \hline
18G                           & 9.0$\pm$3.1             & 120.1$\pm$9.7 & 46.4$\pm$9.0  & 6.2$\pm$5.9  & 2.8$\pm$3.3                                                                              \\ \hline
20G                           & 5.7$\pm$1.5             & 81.1$\pm$5.2  & 49.9$\pm$13.7 & 7.6$\pm$6.7  & 4.4$\pm$4.6                                                                              \\ \hline
25G                           & 7.0$\pm$3.8             & 68.9$\pm$7.1  & 45.9$\pm$9.6  & 6.7$\pm$7.1  & 7.8$\pm$9.7                                                                              \\ \hline
30G                           & 9.3$\pm$4.3             & 88.3$\pm$5.7  & 41.1$\pm$16.6 & 6.4$\pm$6.0  & 6.5$\pm$6.4                                                                              \\ \hline
\end{tabular}
\end{table}

\section{Model capacity}
\begin{figure}[htb]
\centerline{\includegraphics[width=\textwidth]{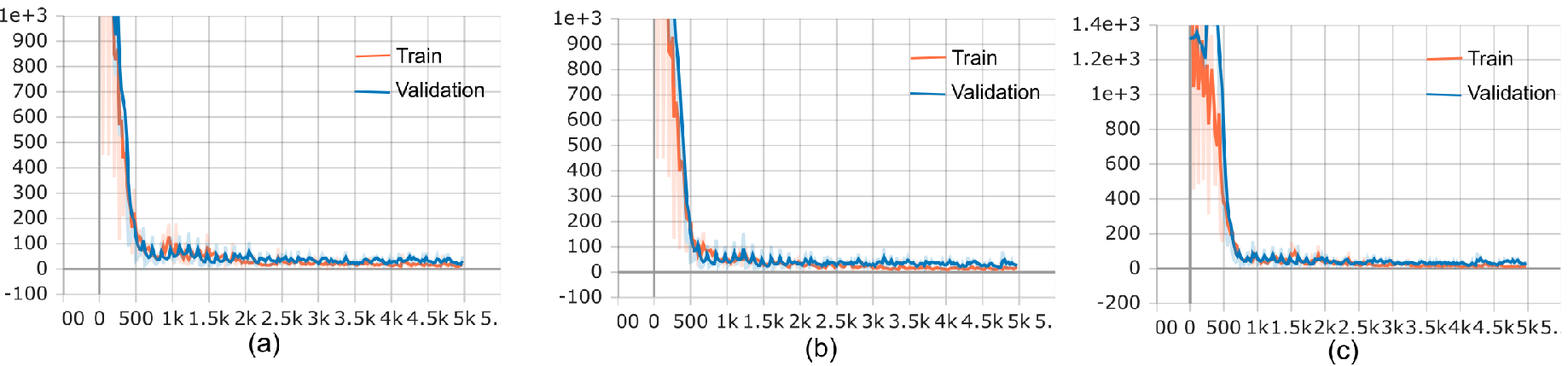}}
\caption{Learning curves of the U-Net with 5 scales (a), 4 scales (b), and 3 scales(c). The generated semi-synthetic dataset, 2000 images with the ground truths, was split into 80\% training set, 10\% validation set, and 10\% testing set. Mean square error (MSE) was reported on the train and validation set per 25 iterations in totaling 5000 iterations.}
\label{supp:model-learning-curves}
\end{figure}

\begin{figure}[htb]
\centerline{\includegraphics[width=0.8\textwidth]{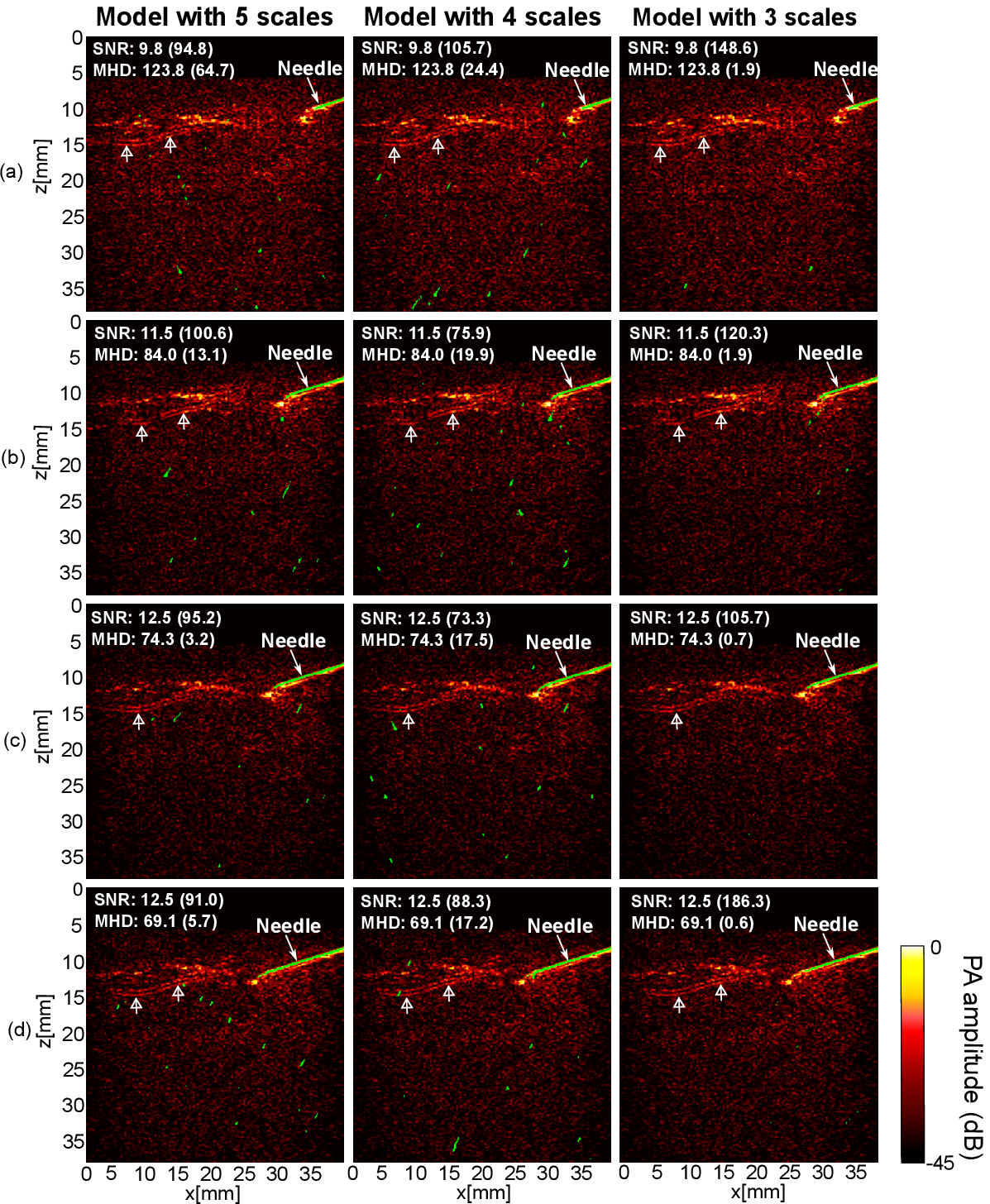}}
\caption{Representative results of model inference on \textit{in vivo} data using the trained model with 5 scales, 4 scales, and 3 scales respectively. The outcomes of the U-Net enhancement are denoted by green lines in photoacoustic (PA) overlays. Signals that may correspond to digital arteries are implied by hollow triangle wide arrows. The signal-to-noise ratio (SNR) and the modified Hausdorff distance (MHD) of the conventional reconstruction (CR) and the U-Net enhancement (in brackets) were measured for each image respectively.  }
\label{supp:model-capacity}
\end{figure}

Figure \ref{supp:model-learning-curves} demonstrates the learning performance of the U-Net with different scales using the generated semi-synthetic dataset. For all the models, the mean square error (MSE) on the train and validation sets both converged to small values after 5000 iterations. Table \ref{tab-capacity} and Figure \ref{supp:model-capacity} further compare their training time, test loss, and inference performance on an \textit{in vivo} PA sequence (128 frames). The proposed model outperformed the models that had larger capacities, indicating its better robustness and generalisability. 

\begin{table}[h]
\centering
\caption {Comparison of training and inferring performance between the U-Net model with different scales. The input of the U-Net models has 128 $\times$ 128 pixels. The inferred image has 256 $\times$ 256 pixels. The test set consisting of 200 images was chosen from the semi-synthetic dataset but blind to the proposed model. The signal-to-noise (SNR) and the modified Hausdorff distance (MHD) were measured on the U-Net enhancement using the inferred data consisting of an \textit{in vivo} sequence of 128 frames.}
\label{tab-capacity}
\begin{tabular}{llll}
\hline
                     & 5 scales       & 4 scales       & 3 scales       \\ \hline
Training time (mins) & 11.10 & 12.13 & 13.43 \\ \hline

MSE on test set                  &45.0$\pm$27.4    & 36.9$\pm$18.8  & 32.5$\pm$13.0
\\ \hline
Inference time (s)   & 0.16           & 0.13           & 0.09           \\ \hline
SNR                  & 95.4$\pm$3.9   & 85.8$\pm$14.8  & 140$\pm$35.6 
\\ \hline
MHD                  & 21.7$\pm$29.0  & 19.8$\pm$3.31  & 4.0$\pm$5.9 
\\ \hline
\end{tabular}
\end{table}

